\documentclass[12pt]{article}
\usepackage{amsmath}
\usepackage{amsfonts}
\usepackage{amssymb}
\usepackage{latexsym}
\usepackage{graphicx}
\usepackage[english]{babel}
\input epsf.sty

\begin{document}

\title {Inflation with Holographic Dark Energy} \vspace{3mm}
\author{{Bin Chen$^{1}$, Miao Li$^{2,3}$, Yi Wang$^{3}$}\\
{\small $^{1}$ School of Physics, Peking University, Beijing 100871, P.R.China}\\
{\small $^{2}$ The Interdisciplinary Center for Theoretical Study}\\
{\small of China (USTC), Hefei, Anhui 230027, P.R.China}\\ {\small
$^{3}$ Institute of Theoretical Physics, Academia Sinica, Beijing
100080, P.R.China}}
\date{}
\maketitle

\begin{abstract}
We investigate the corrections of the holographic dark energy to
inflation paradigm. We study the evolution of the holographic dark
energy in the inflationary universe in detail, and carry out a
model-independent analysis on the holographic dark energy
corrections to the primordial scalar power spectrum. It turns out
that the corrections generically make the spectrum redder. To be
consistent with the experimental data, there must be a upper bound
on the reheating temperature. We also discuss the corrections due
to different choices of the infrared cutoff.
\end{abstract}

\section{Introduction}

It was pointed out in \cite{holoformula} that an ultraviolet (UV)
cutoff of a field theory should be set by an infrared (IR) cutoff
in order that the quantum zero-point energy of a system should not
exceed the mass of a black hole of the same size. Recent studies on the so-called
holographic dark energy are based on this observation. In the original paper
\cite{holoformula}, using the Hubble scale  as the IR cutoff, one
obtains a vacuum energy density comparable to the present day dark
energy. Unfortunately it was later pointed out in \cite{Hsu} that
the resulting equation of state for dark energy does not match the
data. However, one of the present authors (Li) \cite{holoDE} suggested that in cosmological
applications, the IR cutoff should be set by the future event
horizon. This suggestion, combined with the simplest assumption
that dark energy is given by the vacuum energy, provides a
possible solution of the three problems on cosmological
constant \cite{joep}, namely, why the cosmological constant is not
large, nonzero, and comparable to the matter energy density at the
present time.

Following \cite{holoDE}, if one choose the future event horizon as
the IR cutoff, the vacuum energy density is given by
\begin{equation}
\label{HDEdef}
\rho_{\Lambda}=3c^2M_p^2R_h^{-2},
\end{equation}
where $c$ is a constant to be determined by data
fitting of the experiments. Both theoretical reasoning and data fitting suggest $c$ is of
order $1$. $R_h$ is the future event horizon, the boundary of the
volume a fixed observer may eventually observe:
\begin{equation}
R_h\equiv
a\int_t^{\infty}\frac{dt}{a}=a\int_x^{\infty}\frac{dx}{aH},\label{Rh}
\end{equation}
where $x\equiv \ln a$.

In cosmological applications, this vacuum energy density
(\ref{HDEdef}) should couple with gravity and be added to the
energy density term in the Friedmann equation. It  can cause the
late time acceleration of the universe as a component of dark
energy.

As have pointed out by Li \cite{holoDE}, if the holographic dark
energy drives the late time acceleration of the universe, then it
should also affect inflation \cite{inflation}. It is shown that at
roughly 60 e-folds before the end of inflation, the holographic
dark energy becomes comparable with the inflaton energy in
energy density. So the holographic dark energy should induces
corrections to the standard inflation at long wave lengths, these may be
observable at the largest scales of the CMB spectrum. To study how
the holographic dark energy affects inflation and the CMB spectrum
is the main aim of this paper.

This paper is organized as follows. In section 2, we discuss in
more detail the dynamics of the holographic dark energy during
inflation and during the post-inflationary epoch. In section 3, we
study the perturbations and the scalar power spectrum. We perform
a model-independent analysis of the effect of the holographic dark
energy. We find that it leads to a upper bound on the reheating
temperature. Some inflation models such as the $\lambda \varphi^4$
monomial model satisfy this requirement, while other models such
as the $m^2 \varphi^2$ monomial model and the small field
inflation models should be modified to coexist with the
holographic dark energy.

The IR cutoff set by the future event horizon is physically natural
and  agrees empirically with the late time acceleration. However, it
is still unknown in first principle by what mechanism the IR cutoff
sets the UV cutoff and the vacuum energy. Some other kinds of IR
cutoff as well as the linear combinations of them have also been
considered for cosmological purpose\cite{cutoffs}. In section 4, we
consider other IR cutoffs other than the future event horizon,
namely, IR cutoffs set by the particle horizon and the Hubble scale.
We find that for the case of particle horizon, the corrections
during inflation is almost the same as the future event horizon
case. While the IR cutoff set by the Hubble scale seems problematic
since the corrections fail to decay. We end with the conclusions and
discussions in section 5.

\section{The Evolution of Holographic Dark Energy}
For simplicity, we consider inflation driven by a
single minimally coupled inflaton field. The Friedman equation is
\begin{equation}
3M_p^2 H^2=\frac{1}{2}\dot{\varphi}^2+V+3c^2M_p^2R_h^{-2},
\label{Friedman}
\end{equation}

We assume that the holographic dark energy does not couple to the
inflaton. Then the equation of motion of the inflaton is not
affected by the existence of the holographic dark energy,
\begin{equation}
\ddot{\varphi}+3H\dot{\varphi}+V_{\varphi}=0 \label{phieom}.
\end{equation}

>From (\ref{phieom}) and the time derivative of (\ref{Friedman}), we
derive another equation
\begin{equation}
-2M_p^2\dot{H}=\dot{\varphi}^2+2c^2M_p^2R_h^{-2}\left(1-\frac{1}{R_h
H}\right)\label{phidot2}
\end{equation}

And we impose the usual slow-roll conditions,
\begin{eqnarray}
\epsilon&\equiv&-\frac{\dot{H}}{H^2},~~~|\epsilon|\ll 1,\\
\delta&\equiv&-\frac{\ddot{\varphi}}{H\dot\varphi},~~~|\delta|\ll
1,\label{slowroll}
\end{eqnarray}
The validity of these slow-roll conditions will be discussed later
in this section.

In the inflation models without holographic dark energy, the
background equations turns into pure algebraic equations by using
slow roll approximation. While when we take holographic dark
energy into consideration, as $R_h$ is non-local in the scale
factor, we still have to solve a system of differential equations.
In \cite{holoDE}, a convenient method to solve the equations is
carried out by taking
$\Omega_{\Lambda}\equiv\rho_{\Lambda}/(3M_p^2H^2)$ as the unknown
function. The Friedman equation (\ref{Friedman}) is rewritten as
\begin{equation}
\frac{1}{aH}=\frac{1}{a}\sqrt{1-\Omega_{\Lambda}}\sqrt{\frac{3M_p^2}{V}}.\label{FriedmanOL}
\end{equation}
Combining the definition of $R_h$ and $\Omega_{\Lambda}$ with
(\ref{FriedmanOL}), we have the differential equation
\begin{equation}
\Omega_{\Lambda}'=-2\Omega_{\Lambda}(1-\Omega_{\Lambda})\left(1-\frac{\sqrt{\Omega_{\Lambda}}}{c}\right),
\label{OL}
\end{equation}
where the prime denotes the derivative with respect to $x\equiv \ln
a$.

The solution of (\ref{OL}) can be written as
\begin{equation}
-2x+\mbox{constant}=\frac{1}{1-\sqrt{\Omega_{\Lambda}}}+\ln
\Omega_{\Lambda}
-\frac{3}{2}\ln(1-\sqrt{\Omega_{\Lambda}})-\frac{1}{2}\ln(1+\sqrt{\Omega_{\Lambda}}),\label{ceq1solution}
\end{equation}
when $c=1$. In the $\Omega_{\Lambda}\rightarrow0$ limit, the above
equation (\ref{ceq1solution}) reduces to $\Omega_{\Lambda}\sim
a^{-2}$, in agreement with \cite{holoDE}.

For the $c\neq 1$ case,
\begin{eqnarray}
-2x+\mbox{constant}&=&-\frac{1}{1-c^2}\left(- c^2\ln
   (1-\Omega_{\Lambda})+ c^2\ln \Omega_{\Lambda} +c
   \ln \frac{1+\sqrt{\Omega_{\Lambda}}}{1-\sqrt{\Omega_{\Lambda}}}
   \right. \nonumber\\ &&\left.+2 \ln
   \left(c-\sqrt{\Omega_{\Lambda}}\right)-\ln
   \Omega_{\Lambda}\right)
   \label{cneq1solution}
\end{eqnarray}
The $c\neq 1$ case has a smooth $c=1$ limit, which agrees with
(\ref{ceq1solution}). A divergence occurs at $\Omega_{\Lambda}=c^2$.
It is due to the big rip singularity in the phantom-like cosmology when
$c<1$.
It can be shown that when $\Omega_{\Lambda}\rightarrow0$, it also
scales as $\Omega_{\Lambda}\sim a^{-2}$, the same as in the $c=1$
case.

\begin{figure}
\centering
\includegraphics[totalheight=2.5in]{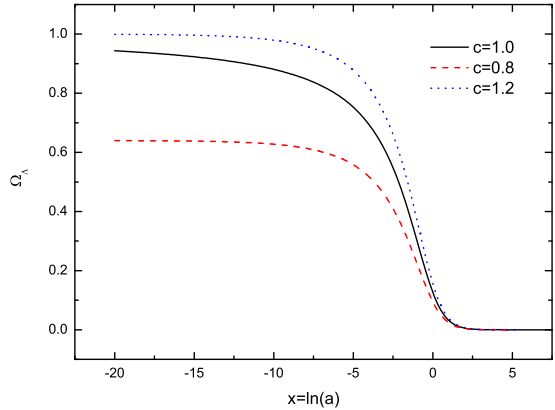}
\caption{\small{The evolution of $\Omega_{\Lambda}$ as a function of
the logarithm of the scale factor $a$. Here $a$ is not normalized to
$a_0=1$, the normalization is chosen for numerical convenience. We
neglected the change of the inflaton energy density in time for
simplicity. }} \label{fig:bgevolution}
\end{figure}

We present the numerical solutions in Fig.\ref{fig:bgevolution},
where we give one example in each class ($c=1$, $c<1$ and $c>1$).  Note that
$c=1$ is a natural value of $c$, $c=0.8$ is the best fit value for
SN+CMB+BAO\cite{holodata}, and $c=1.2$ is taken for comparison. In
\cite{holodata}, there are best fit values from other experiments
such as $c=0.6$ from X-ray gas mass fraction of rich clusters, the solution
behaves qualitatively the same  in each class. We observe
that when $c\geq1$, the universe will undergo a long holographic
dark energy dominated period given a dark energy dominated initial
condition. While once the inflaton takes a significant part in the
energy density, it will quickly become dominating. In the $c<1$ case,
the critical point lies at $\Omega_{\Lambda}=c^2$ instead of
$\Omega_{\Lambda}=1$. It is due to the divergence at
$\Omega_{\Lambda}=c^2$ in (\ref{cneq1solution}). Note that in this
figure we have assumed the inflaton energy stays constant. A more
rigorous calculation should take the time variation of $\varphi$
into account.

Other background quantities such as $H$ and $\varphi$ during
inflation can be calculated using $\Omega_{\Lambda}$ by
(\ref{FriedmanOL}) and (\ref{phieom}).

Since it seems difficult to give an initial condition of
$\Omega_{\Lambda}$ from first principle, we determine
$\Omega_{\Lambda}$ by experiments. From the WMAP
experiment\cite{WMAP3yr}, $\Omega_{\Lambda}\simeq0.76$ in the
present time. This can help to fix an initial condition for
$\Omega_{\Lambda}$ during inflation.

It can be shown that in the radiation dominated and matter dominated
epochs, the scaling property of $\rho_{\Lambda}$ is
$\rho_{\Lambda}\sim a^{-2}$. Then, from the number of
e-folds\cite{RiottoLyth},
\begin{equation}
N_{\rm{COBE}}=62-\ln(10^{16}\rm{GeV}/V^{1/4}_{\rm{end}})-\frac{1}{3}\ln
(V^{1/4}_{\rm{end}}/\rho^{1/4}_{\rm{reh}}),
\end{equation}
when we assume that the reheating happens immediately after the
inflation, and the reheating period is short enough, we get the
initial value of $\rho_{\Lambda}$,
\begin{equation}
\rho_{\Lambda
COBE}=1.2\times10^{-9}\left(\frac{T_{\rm{reh}}}{10^{16}\rm{GeV}}\right)^4M_p^4=4.2
T_{\rm{reh}}^4
\end{equation}
Translated into $R_h$, this is
\begin{equation}
R_h=5c\times
10^4\left(\frac{10^{16}\rm{GeV}}{T_{\rm{reh}}}\right)^2
M_p^{-1}.\label{Rhnum}
\end{equation}
It is very sensitive to $T_{\rm{reh}}$, the reheating temperature.
We will see that the above relation provides a constraint on the
reheating temperature, after combined with the effect of the
holographic dark energy on the spectral index.

Before calculating the perturbations, let us pause to discuss the
validity of the slow-roll conditions. From (\ref{OL}) we  get
\begin{equation}
\dot{\Omega}_{\Lambda}=2H\Omega_{\Lambda}\left(\frac{\sqrt{\Omega_{\Lambda}}}{c}-1\right)
(1-\Omega_{\Lambda}).
\end{equation}
So $\Omega_{\Lambda}$ is a slow-roll quantity only when
$\Omega_{\Lambda}\rightarrow 1$. In the $c<1$ case,
$\Omega_{\Lambda}\rightarrow c^2$ can satisfy the slow-roll
condition as well. Then from Friedman equation in flat space
\begin{equation}
\Omega_{\varphi}+\Omega_{\Lambda}=1,
\end{equation}
we conclude that $\Omega_{\varphi}$ is a slow-roll quantity only
when $\Omega_{\Lambda}\rightarrow0$ or
$\Omega_{\Lambda}\rightarrow1$ (and $\Omega_{\Lambda}\rightarrow
c^2$ when $c<1$). For a intermediate value of $\Omega_{\Lambda}$, either $\epsilon$ or
$\delta$ slow-roll condition (or both of them) is broken. In the
following, we mainly consider the limit when $\Omega_{\Lambda}$ is
small enough not to spoil the slow-roll conditions. Although the
slow-roll regime does not take a large part in the whole parameter
space, it is interesting enough from the experimental point of
view. It is because as we will see in the following sections, the
existence of the holographic dark energy produces a rather red
spectrum. A slow-roll inflation with a small part of holographic
dark energy already conflicts with the CMB experiment if this
regime lies in the observable part of inflation. What we are
interested in is to give a constraint of the upper bound of
holographic dark energy at the beginning of the observable
inflation. This constraint can be translated into a constraint on
the reheating temperature. So to study the slow-roll regime is
enough for our purpose.

\section{Perturbations and the Spectrum}

Since the holographic dark energy depends only on the background
quantities, it does not seem to produce its own perturbations. So
the standard perturbation equations \cite{MFB92} stay unchanged.
>From these equations, and use the equation (\ref{phieom}), we can
get the diagonized equation for $\phi$ in the longitudinal gauge,
\begin{equation}
\ddot{\phi}+\left(H-\frac{2\ddot{\varphi}}{\dot{\varphi}}\right)\dot{\phi}+\left(4\dot{H}
-H\frac{2\ddot{\varphi}}{\dot{\varphi}}+\frac{\dot{\varphi}^2}{M_p^2}\right)\phi-\frac{\nabla^2}{a^2}\phi=0,
\label{purteom}
\end{equation}
where $\phi$ is the perturbation of the metric,
$
ds^2=a^2\left(-(1+2\phi)d\tau^2+(1-2\phi)dx^idx^i\right).
$

Let $u\equiv \phi/\dot{\varphi}$, then the equation for $u$ reads
\begin{equation}
\frac{d^2}{d\tau^2}u_k+\frac{1}{\tau^2}\left(-4\epsilon+\delta+\sigma\right)u_k+k^2
u_k=0, \label{bessel}
\end{equation}
where $\tau$ is the conformal time and
\begin{equation}
\sigma\equiv\frac{\dot{\varphi}^2}{H^2 M_p^2}.
\end{equation}

It takes the form of a Bessel equation. As usual, we can write down
its solution and determine the normalization constants by the usual
procedure.

In order to compare with the CMB experiments, it is useful to
single out a conserved quantity after horizon crossing. The usual
comoving curvature perturbation is not conserved when the
holographic dark energy is present during inflation. It is because
the inflaton field fluctuates while the holographic dark energy
does not, then the perturbation is not adiabatic. In this case, we
can use a nearly conserved quantity instead (for a derivation of
the general nearly conserved quantity in a slow-roll inflation,
see \cite{nmi}),
\begin{equation}
\mathcal{R}=\frac{2M_p^2H^2}{\dot{\varphi}^2}\left(\frac{\dot{\phi}}{H}+\phi\right)\exp
\left(2c^2\int_t^{t_{LS}}\frac{1}{R_h^2H}\left(1-\frac{1}{R_hH}\right)dt\right),
\end{equation}
with the power spectrum
\begin{equation}
\mathcal{P}_{\mathcal{R}}=\frac{H^4}{4\pi^2\dot{\varphi}^2} \exp
\left(4c^2\int_t^{t_{LS}}\frac{1}{R_h^2H}\left(1-\frac{1}{R_hH}\right)dt\right),
\end{equation}
where $t_{LS}$ is the time of the last scattering surface. Since the
holographic dark energy decays fast, where to put the upper bound of
the integration is not important.

The spectral index takes the form
\begin{equation}
n_s-1=-8\epsilon+2\delta+2\sigma=-4\epsilon+2\delta-\frac{4c^2}{R_h^2H^2}\left(1-\frac{1}{R_hH}\right).
\end{equation}

Now let us derive a model-independent formula for the corrections
to the spectral index produced by the holographic dark energy.
Since the slow roll parameter
\begin{equation}\eta\equiv\epsilon+\delta\end{equation}
 depends only on the
form of the inflaton potential and the Hubble constant, it is not
influenced by the holographic dark energy. The inflaton potential
is set by the model, so is not affected by the existence of the
holographic dark energy. The Hubble constant is calculated
backwards $N_{\rm{COBE}}$ e-folds from the final condition of
inflation. As we will see, at the time later than that of
$N_{\rm{COBE}}$ during inflation, the holographic dark energy
should not affect the evolution of the universe. So the Hubble
constant at $N_{\rm{COBE}}$ is also not changed whether the
holographic dark energy exists or not. When $\Omega_{\Lambda}$ is
small, the slow roll parameter $\epsilon$ can be decomposed into

\begin{equation}
\epsilon=-\frac{1}{2}
\sqrt{\frac{3M_p^2}{V^3}}V_{\varphi}\dot{\varphi}
+\frac{c^2}{R_h^2H^2}\left(1-\frac{1}{R_hH}\right)
\equiv\epsilon_0+\frac{c^2}{R_h^2H^2}\left(1-\frac{1}{R_hH}\right)
\end{equation}
where $\epsilon_0$ is the original contribution in the inflation
models without the holographic dark energy, and
$\frac{c^2}{R_h^2H^2}\left(1-\frac{1}{R_hH}\right)$ is a
correction term.

\begin{figure}
\centering
\includegraphics[totalheight=2.5in]{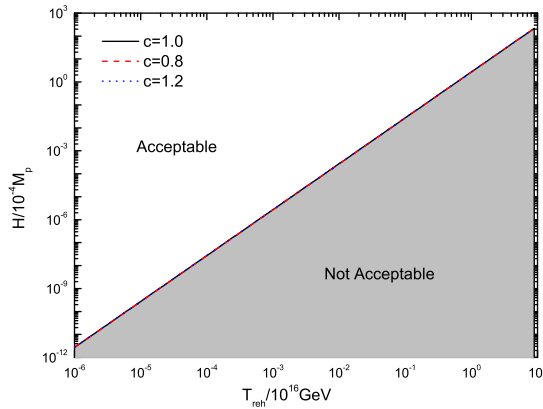}
\caption{\small{Constraint on the Hubble scale during inflation and
the reheating temperature. The lower half of the figure is not
favored by experiments because it produces too red a spectrum (the
correction to $n_s-1$ is smaller than $-0.05$). The correction is
not sensitive to $c$.}} \label{fig:Treh}
\end{figure}

The spectral index can be written as
\begin{equation}
n_s-1=-6\epsilon_0+2\eta-\frac{10c^2}{R_h^2H^2}\left(1-\frac{1}{R_hH}\right)
\end{equation}
where $-6\epsilon_0+2\eta$ is the standard contribution from the
single field inflaton models without the holographic dark energy.
Therefore, we see that the effect of the holographic dark energy
is to make the spectrum redder. Note that from experiment we do
not expect too red a spectrum, there is a constraint from the
experiment on $R_hH$. The constraint can be translated into a
constraint on the Hubble constant during the inflation and the
reheating temperature by (\ref{Rhnum}). A constraint
$-\frac{10c^2}{R_h^2H^2}\left(1-\frac{1}{R_hH}\right)>-0.05$ is
shown in Fig.\ref{fig:Treh}. If we suggest further that
$T_{\rm{reh}}\simeq V_{\rm{end}}^{1/4}$, then the constraint on
the reheating temperature is
\begin{equation}
T_{\rm{reh}}^4<1.2\times 10^{-3}V_{\rm{COBE}}\label{Tconstraint}
\end{equation}
The equation (\ref{Tconstraint}) tells us that the energy scale
during the last $N_{\rm{COBE}}$ e-folds of inflation should drop
three orders of magnitude. This provides a constraint on the
inflation models flavored by the holographic dark energy.

For the monomial potential $V=\lambda M_p^{4-p}\varphi^p$, if we
assume inflation ends at roughly $\epsilon=1$, $\varphi$ can be
expressed as a function of the e-folding number as
\begin{equation}
\varphi_N=\sqrt{2pN+\frac{p^2}{2}}M_p
\end{equation}
so the relation between $T_{\rm{reh}}$ and $V_{\rm{COBE}}$ can be
expressed as
\begin{equation}
T_{\rm{reh}}^4=\left(\frac{4N_{\rm{COBE}}}{p}+1\right)^{-\frac{p}{2}}V_{\rm{COBE}}
\end{equation}

For example, the $m^2\varphi^2$ model gives
$T_{\rm{reh}}^4\simeq8\times 10^{-3}V_{\rm{COBE}}$, so for this
model other mechanism is needed to be consistent with the existence
of the holographic dark energy. While the $\lambda\varphi^4$ model
has $T_{\rm{reh}}^4\simeq3\times 10^{-4}V_{\rm{COBE}}$, which
is consistent with the existence of the holographic dark
energy.

For the small field inflation models,
$V=V_0\left(1-\left(\frac{\varphi}{M}\right)^p\right),p\geq2$, the
energy scales do not change much during inflation. So the small
field inflation models are not favored by the existence of the
holographic dark energy.

We do not need to worry about the late time entropy perturbation,
because the entropy perturbation decays fast with the decay of the
holographic dark energy.

It is also an interesting issue whether and how the IR cutoff set by
the future event horizon affect the observed CMB spectrum. In
\cite{cutoffCMB}, it has been studied in a simple toy model how the
finiteness of the future event horizon influences the angular power
spectrum, especially the low $l$ behavior. Since the boundary
conditions studied in \cite{cutoffCMB} are not respected during the
expansion of the universe, we take a slightly different treatment:
we simply cut off the primordial spectrum and set the primordial
spectrum to zero when $\pi/k$ is larger than the present future
event horizon. When $\pi/k$ is smaller than the present future event
horizon, the primordial spectrum is not modified.

\begin{figure}
\centering
\includegraphics[totalheight=2.5in]{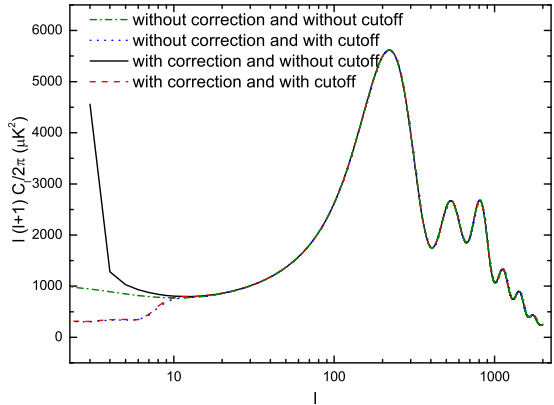}
\caption{\small{The CMB temperature angular power spectrum. In
this figure, we considered the spectrum with and without
primordial holographic dark energy corrections, as well as with
and without late time direct IR cutoff. If we take the late time
IR cutoff into consideration, the curves with and without
holographic dark energy nearly coincide. So we see that the late
time IR cutoff can hide the corrections given by the primordial
holographic dark energy. And without the late time IR cutoff, the
power spectrum can become extremely red in some parameter region.
This together with Fig.\ref{fig:Treh} provide a constraint on the
Hubble scale during inflation and the reheating temperature.}}
\label{fig:CMB}
\end{figure}

For illustrative purpose, we show in Fig.\ref{fig:CMB} the CMB
 power spectrum with and
without the contribution of holographic dark energy, as well as
with and without the late time IR cutoff. The parameter for the
holographic dark energy sector is chosen as the best fit value of
SN+CMB+BAO\cite{holodata}. In Fig.\ref{fig:CMB}, we take a large
correction from holographic dark energy as an example. Such a
large correction conflicts with the experimental data, and as
shown in Fig.\ref{fig:Treh}, it is in the ``Not Acceptable"
region.

Now we have seen that even a small amount of holographic dark
energy can produce a rather red spectrum, which can give a
constraint on the reheating temperature $T_{\rm{reh}}$. To have a
better understanding of the whole picture, here we also present a
simple (and crude) estimate for the regime where the holographic
dark energy is dominate or of the same order as the energy density
of the inflaton. As discussed in the last section, in this regime,
the slow-roll approximation breaks down. According to
\cite{fastroll}, even for the fast rolling we can still take
\begin{equation}
\delta\varphi\sim\frac{H}{2\pi},
\end{equation}
just like the slow-roll inflation case, where $\delta\varphi$ is the
frozen amplitude of the inflaton field fluctuations outside the
horizon. The existence of the holographic dark energy results in a
larger Hubble constant, so the fluctuations generally becomes larger
in this regime. This implies that the power spectrum was made redder
by a large holographic dark energy in the early stage of the
inflation. Therefore, the observed fluctuations of the inflaton
should be created during the slow-roll regime.

\section{Other Realizations of Holographic Principle}

Although the IR cutoff set by the future event horizon is
theoretically natural and experimentally interesting, there are
other candidates of IR cutoffs. The arbitrariness comes from the
lack of knowledge of the holographic principle. So in this
section, we consider as well two other kinds of IR cutoffs,
namely, the particle horizon and the Hubble scale, even though for
these two cases the equation of state for dark energy does not
agree with data. Note that the perturbation equations
(\ref{purteom}) and (\ref{bessel}) still hold for these two cases.

For the particle horizon case, the formulae are similar to the case
with the future event horizon. The particle horizon is defined as
\begin{equation}
R_H=a\int_0^t\frac{dt}{a},
\end{equation}
and the vacuum energy density is given by
\begin{equation}
\rho_{\Lambda}=3 c^2 M_p^2 R_H^{-2}.
\end{equation}

The Friedman equation takes the form
\begin{equation}
3M_p^2 H^2=\frac{1}{2}\dot{\varphi}^2+V+3c^2M_p^2R_H^{-2},
\label{PHFriedman}
\end{equation}

Take derivative respect to the equation (\ref{PHFriedman}), and use
(\ref{phieom}), we can get
\begin{equation}
-2M_p^2\dot{H}=\dot{\varphi}^2+2c^2M_p^2R_H^{-2}\left(1+\frac{1}{R_H
H}\right)\label{PHphidot2}
\end{equation}

For the background equations expressed by $\Omega_{\Lambda}$, we
have
\begin{equation}
\Omega_{\Lambda}'=-2\Omega_{\Lambda}(1-\Omega_{\Lambda})\left(1+\frac{\sqrt{\Omega_{\Lambda}}}{c}\right)
\end{equation}
where $\Omega_{\Lambda}$ is the ratio of the vacuum energy density
to the critical density in the particle horizon case.

As in the future event horizon case, we can get the nearly conserved
quantity,
\begin{equation}
\mathcal{R}=\frac{2M_p^2H^2}{\dot{\varphi}^2}\left(\frac{\dot{\phi}}{H}+\phi\right)\exp
\left(2c^2\int_t^{t_{LS}}\frac{1}{R_H^2H}\left(1+\frac{1}{R_HH}\right)dt\right),
\end{equation}
with the power spectrum
\begin{equation}
\mathcal{P}_{\mathcal{R}}=\frac{H^4}{4\pi^2\dot{\varphi}^2} \exp
\left(4c^2\int_t^{t_{LS}}\frac{1}{R_H^2H}\left(1+\frac{1}{R_HH}\right)dt\right),
\end{equation}
and the spectral index
\begin{equation}
n_s-1=-8\epsilon+2\delta+2\sigma=-4\epsilon+2\delta-\frac{4c^2}{R_H^2H^2}\left(1+\frac{1}{R_HH}\right).
\end{equation}

The spectral index can be split into two parts each related to the usual slow-roll
inflation and to the holographic dark energy
\begin{equation}
n_s-1=-6\epsilon_0+2\eta-\frac{10c^2}{R_H^2H^2}\left(1+\frac{1}{R_HH}\right)
\end{equation}
Obviously the presence of the holographic dark energy makes the
spectrum redder.

The vacuum energy density during inflation can be calculated
similarly to the future event horizon case. If we assume that the vacuum
energy related to the particle horizon is of the same order compared
with the matter density in the present time, then we have at the
e-folds $N_{\rm{COBE}}$ before the end of inflation, $R_H\sim
R_h$. So the IR cutoff set by $R_H$ gives a similar, but slightly
redder spectrum than the future event horizon case.

Finally, let's consider the IR cutoff set by the Hubble scale. The vacuum energy density is set to
\begin{equation}
\rho_{\Lambda}=3 c^2 M_p^2 H^{2}.
\end{equation}

The Friedman equation can be written as,
\begin{equation}
3M_p^2(1-c^2) H^2=\frac{1}{2}\dot{\varphi}^2+V \label{HFriedman}
\end{equation}
and as usual we can get the following equation on
$\dot{\varphi}^2$
\begin{equation}
-2M_p^2(1-c^2)\dot{H}=\dot{\varphi}^2 \label{Hphidot2}
\end{equation}

Although $c=1$ may be the most natural value of $c$, we note that
$c=1$ forces both the kinetic and potential energy of $\varphi$
vanish. A small departure from $c=1$ leads to a smaller required
value of $\varphi$ for a given Hubble constant during inflation.
It is not surprising since the IR cutoff set by $H$ can be think
of as a redefinition of the Planck mass, and a smaller effective
Planck mass (i.e. a stronger gravitational coupling) leads to a
smaller required value of the inflaton to drive inflation.
However, when we consider the perturbations and calculate the
spectrum, the difference is not simply a rescaling of Planck mass.
This is because just as in the future event horizon case, the
vacuum energy produced by holographic principle depends nonlocally
on the Hubble constant, so the vacuum energy do not produce
perturbation by itself.

The nearly conserved quantity takes the form
\begin{equation}
\mathcal{R}=\frac{2M_p^{2}H^{2+2c^2}}{M^{-2c^2}\dot{\varphi}^2}\left(\frac{\dot{\phi}}{H}+\phi\right),
\end{equation}
with the power spectrum
\begin{equation}
\mathcal{P}_{\mathcal{R}}=\frac{H^{4+4c^2}}{4\pi^2M^{4c^2}\dot{\varphi}^2},
\end{equation}
where $M$ is a constant of the dimension of energy.

The spectral index can be written as
\begin{equation}
n_s-1=-8\epsilon+2\delta+2\sigma=-(4+4c^2)\epsilon+2\delta.
\end{equation}

Note that the power spectrum is suppressed from the usual one by a
factor $(H/M)^{4c^2}$. If we take $M$ as the Planck scale or the
string scale, the problem of a hierarchy of scales during inflation
can be solved or eased by this kind of IR cutoff. But $M$ can not be
determined as usual by the final condition of inflation. Because the
models with this kind of IR cutoff have a non-decaying correction
from the holographic dark energy. As a toy model, it can not come
back to the standard inflation models.

Therefore,  to take the Hubble scale  as the IR cutoff is
problematic. The long-lived entropy perturbation is produced. We
need extra mechanism for the entropy perturbation to decay. And
furthermore, the post inflationary period is problematic because
the non-decaying large correction to the standard picture may
modify some exactly known results such as the nucleosynthesis. One
possible solution is to assume that the holographic bound for the
vacuum energy is not saturated for a large portion of the history
of the universe. In \cite{nonsaturated}, such a possibility is
investigated.

\section{Conclusions}

In conclusion, in this paper, a detailed study on the effects of
the holographic dark energy to the inflation is carried out.

We find that the corrections on the spectral index generally make
the scalar primordial power spectrum redder than that without the
holographic dark energy. In order that the perturbations do not
become too large to conflict with the experiments, a constraint on
the reheating temperature is given. The constraint is that the
energy scale  should drop three orders of magnitude during the
last $N_{\rm{COBE}}$ e-folds of inflation. As we take the monomial
potential for inflation for example, the $\lambda\varphi^4$ model
satisfies the requirement while the monomial $m^2\varphi^2$ model
and the small field inflation models have to add extra mechanism
to live with the holographic dark energy. In our discussion, we
assume that the reheating happens right after the end of
inflation, it could be interesting to consider the case of
late-time reheating.

We compared the IR cutoff set by the particle horizon and the
Hubble scale during inflation with that by the future event
horizon. We find that the cutoff set by the particle horizon
provide a similar but slightly redder power spectrum than the one
produced in the future event horizon case. The cutoff set by the
Hubble scale seems problematic since the correction does not decay
during the whole evolutionary history of the universe, which may
spoil some exactly known results.
%We also considered the late time
%IR cutoff brought forward by \cite{cutoffCMB}. We find that this
%IR cutoff can hide the corrections in the early inflationary
%universe.

\section*{Acknowledgments}
This work was supported by grants of NSFC. BC was also supported by
the Key Grant Project of Chinese Ministry of Education (NO. 305001).
We thank Xin Zhang for discussions.

\end{document}